\documentstyle[psfig]{aipproc}

\def\la{\mathrel{\hbox{\rlap{\hbox{\lower4pt\hbox{$\sim$}}}\hbox{$<$}}}}
\def\ga{\mathrel{\hbox{\rlap{\hbox{\lower4pt\hbox{$\sim$}}}\hbox{$>$}}}}

\begin{document}
\title{The GRB/SN Connection:  An Improved Spectral Flux Distribution for the Supernova Candidate Associated with GRB 970228}

\author{Daniel E. Reichart$^*$, Francisco J. Castander$^{\dagger}$, and
Donald Q. Lamb$^*$} \address{$^*$Department of Astronomy \&
Astrophysics, University of Chicago,\\ 5640 South Ellis Avenue, Chicago
IL, 60637\\ $^{\dagger}$Observatoire Midi-Pyr\'en\'ees, 14 Av. Edouard
Belin, 31400 Tolouse, France}

\maketitle

\begin{abstract}    
We better determine the spectral flux distribution of the supernova
candidate associated with GRB 970228 by modeling the spectral flux
distribution of the host galaxy of this burst, fitting this model to
measurements of the host galaxy, and using the fitted model to better
subtract out the contribution of the host galaxy to measurements of the
afterglow of this burst.      
\end{abstract}

\section*{Introduction} 

The discovery of what appear to be SNe dominating the light curves and
spectral flux distributions (SFDs) of the afterglows of GRB 980326
(Bloom et al. 1999) and GRB 970228 (Reichart 1999; Galama et al. 1999)
at late times after these bursts strongly suggests that at least some,
and perhaps all, of the long bursts are related to the deaths of
massive stars.  Here, we build upon the results of Reichart (1999) by
modeling the SFD of the host galaxy of GRB 970228, fitting this
model to measurements of the host galaxy, and using the fitted model
to better subtract out the contribution of the host galaxy to
measurements of the afterglow of this burst.

\section*{Observed and Modeled SFDs for the Host Galaxy} 

In Figure 1, we plot the observed SFD of the host galaxy of GRB 970228,
as measured with {\it HST}/WFPC2, {\it HST}/NICMOS2, and Keck I
(Castander \& Lamb 1999a; Fruchter et al. 1999).  To these measurements
and a broadband measurement made with {\it HST}/STIS (Castander \&
Lamb 1999a; Fruchter et al. 1999), which is not plotted, we fit a
two-parameter, spectral synthesis model (see Castander \& Lamb 1999a for
details).  The two parameters are the normalization of the SFD, and the
age of the galaxy, defined to be the length of time that star formation
has been occurring at a constant rate.  Taking
$A_V = 1.09$ mag for the Galactic extinction along the line of sight
(Castander \& Lamb 1999b), we find a fitted age of $270^{+460}_{-180}$
Myr; different values of $A_V$ affect primarily the fitted age, and not
the fitted SFD.  Furthermore, models in which star formation slows
considerably, or ceases, are generally too red to account
for the measurements.  Finally, we note that the fitted J- and R-band
spectral fluxes are perfectly consistent with what one finds simply
from linear interpolation between adjacent photometric bands. 

\begin{figure}[tb]
\psfig{file=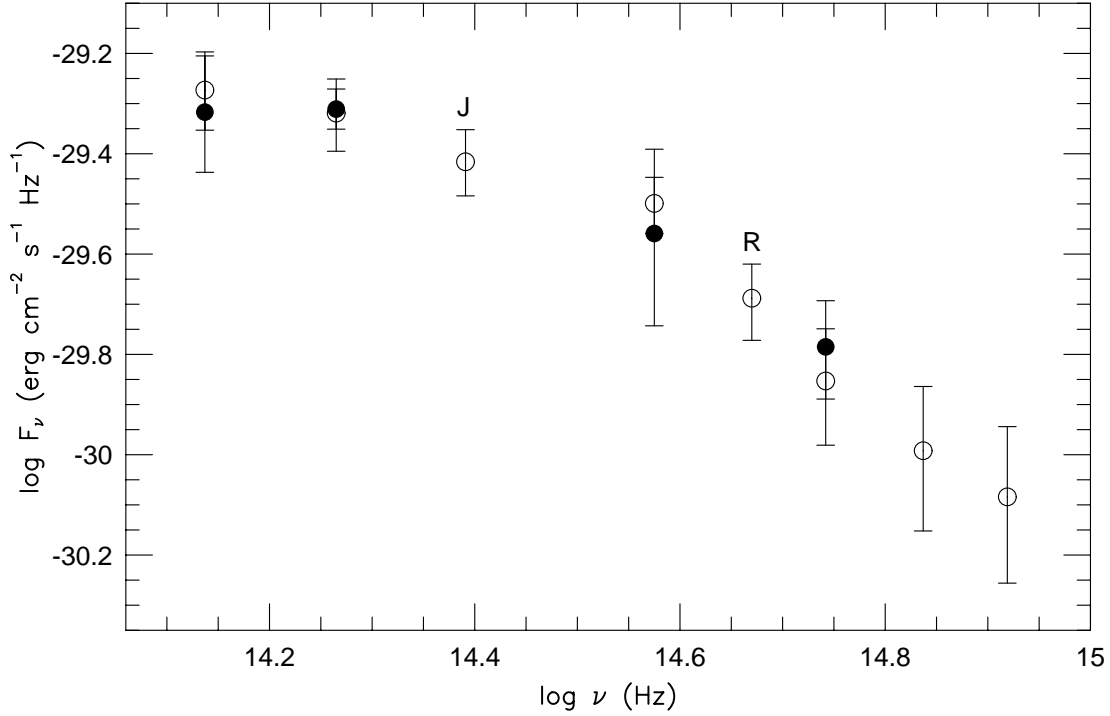,width=5.75truein,clip=}
\caption{The observed (filled circles) and modeled (unfilled circles) K- through U-band SFDs of the host galaxy of GRB 970228.}
\end{figure}

\section*{SFD of the SN Candidate Derived using the Observed SFD of the Host Galaxy} 

In Figure 2, we plot the SFD of the afterglow minus the {\it observed}
SFD of the host galaxy from Figure 1.  For the SFD of the afterglow, we
use the revised K-, J-, and R-band measurements of Galama et al. (1999)
and the I- and V-band measurements of Castander \& Lamb (1999a; see also
Fruchter et al. 1999); all of these measurements were taken between 30
and 38 days after the burst.  We have scaled these measurements to a
common time of 35 days after the burst, and have corrected these
measurements for Galactic extinction along the line of sight (see
Reichart 1999 for details).  The K-band measurement of the afterglow is
consistent with that of the host galaxy (Galama et al. 1999), resulting
in an upper limit in Figure 2; J- and R-band measurements of the host
galaxy are not available, again resulting in upper limits in Figure 2. 
As originally concluded by Reichart (1999), this SFD is consistent with
that of SN 1998bw, after transforming it to the redshift of the burst,
$z = 0.695$ (Djorgovski et al. 1999), and correcting it for Galactic
extinction along its line of sight (see Reichart 1999 for details).

\begin{figure}[t]
\psfig{file=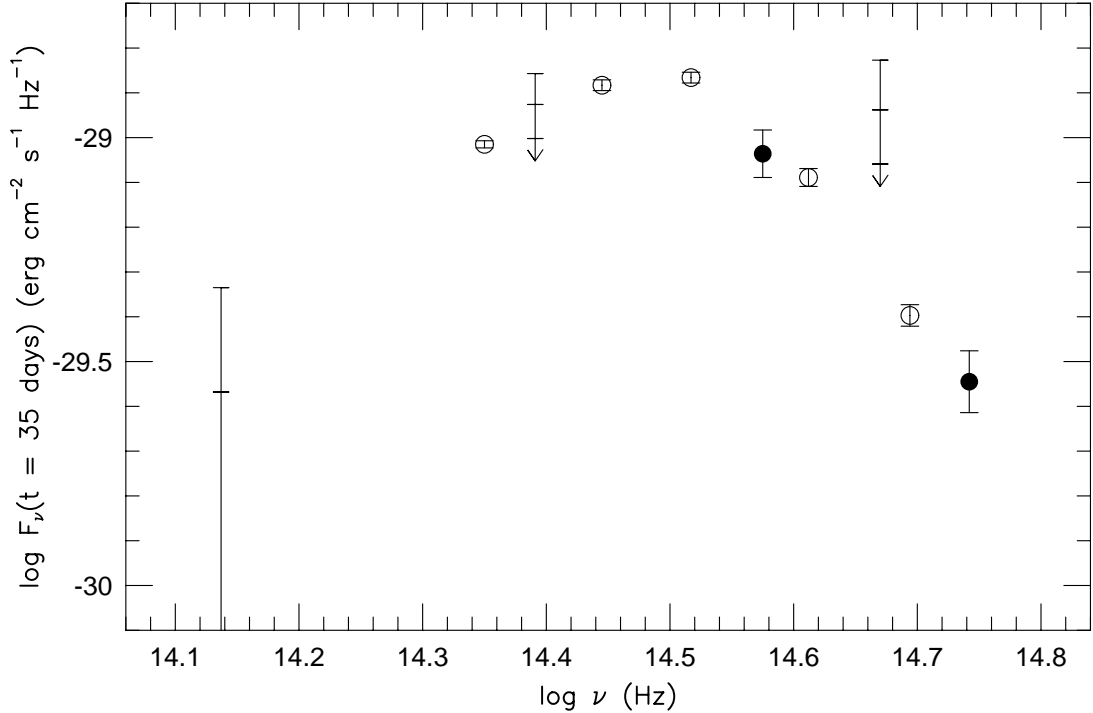,width=5.75truein,clip=} 
\caption{The K- through V-band SFD of the late afterglow of GRB 970228 after subtracting out the {\it observed} SFD of the host galaxy from Figure 1 and correcting for Galactic extinction (filled circles and upper limits), and the I- through U-band SFD of SN 1998bw after transforming to the redshift of GRB 970228, $z = 0.695$, and correcting for Galactic extinction (unfilled circles).  The K-, J-, and R-band upper limits are 1, 2, and 3 $\sigma$.}
\end{figure}

\section*{SFD of the SN Candidate Derived using the Modeled SFD of the Host Galaxy} 

In Figure 3, we plot the same distribution, but minus the {\it modeled}
SFD of the host galaxy from Figure 1.  The SN-like component to the
afterglow is detected in the J band, and possibly in the R band.  The
J-band measurement suggests that the SN-like component is $\approx 1/2$
mag fainter, and $\approx 1/2$ of a photometric band bluer, than SN
1998bw; however, this difference in J-band spectral fluxes is
significant only at the $\approx 2.5$ $\sigma$ level.  When possible
photometric zero point errors and uncertainties in our spectral
synthesis model of the SFD of the host galaxy are included, this
difference is significant only at the $\approx 2$ $\sigma$ level. 
However, it is suggestive of what is generally expected:  the Type Ic
SNe that are theorized to be associated with bursts (e.g., Woosley
1993) are not expected to be standard candles.

\begin{figure}[t]
\psfig{file=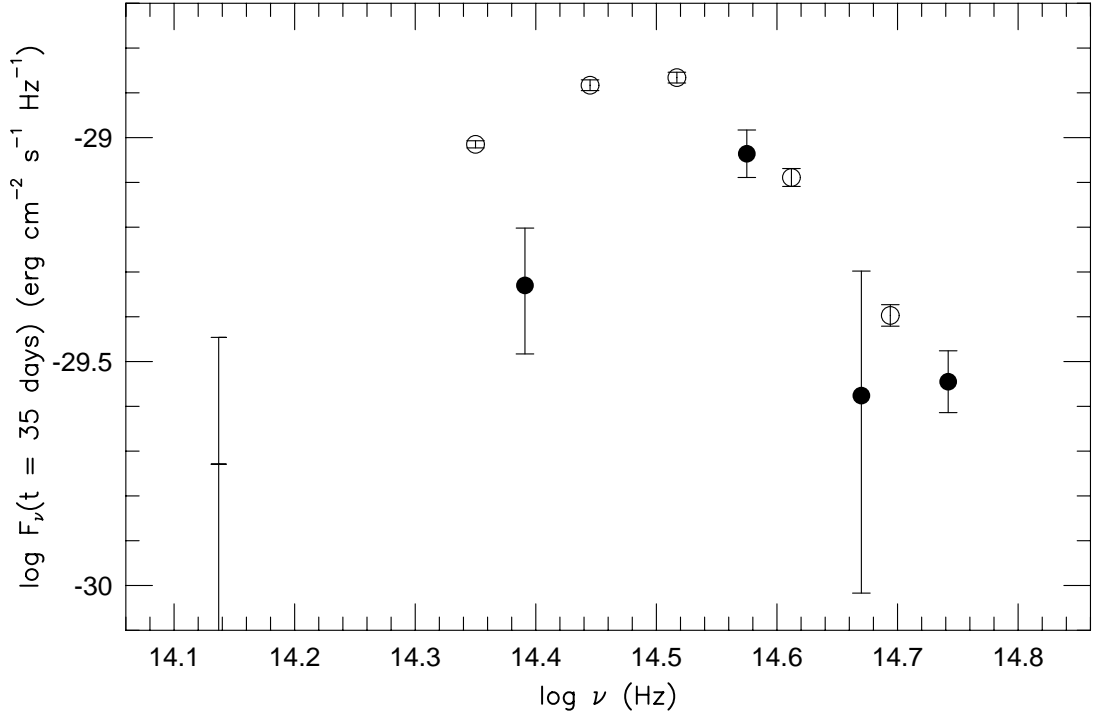,width=5.75truein,clip=}
\caption{The K- through V-band SFD of the late afterglow of GRB 970228 after subtracting out the {\it modeled} SFD of the host galaxy from Figure 1 and correcting for Galactic extinction (filled circles and upper limits), and the I- through U-band SFD of SN 1998bw after transforming to the redshift of GRB 970228, $z = 0.695$, and correcting for Galactic extinction (unfilled circles).  The K-band upper limits are 1, 2, and 3 $\sigma$.}
\end{figure}


\begin{references}
\bibitem{bea99}
Bloom, J. S., et al. 1999, Nature, 401, 453 
\bibitem{cl99a}
Castander, F. J., \& Lamb, D. Q. 1999a, ApJ, 523, 593 
\bibitem{cl99b}
Castander, F. J., \& Lamb, D. Q. 1999b, ApJ, 523, 602
\bibitem{dea99}
Djorgovski, S. G., et al. 1999, GCN Report 289 
\bibitem{fea99}
Fruchter, A. S., et al. 1999, ApJ, 516, 683 
\bibitem{gea99}
Galama, T. J., et al. 1999, ApJ, submitted 
\bibitem{r99}
Reichart, D. E., 1999, ApJ, 521, L111 
\bibitem{w93}
Woosley, S. E. 1993, ApJ, 405, 273 
\end{references}
\end{document}